\documentclass[12pt]{article}
\usepackage{amsmath,amsthm,amssymb,euscript,epsfig,youngtab}
\usepackage{array}
\usepackage{fancybox}
\usepackage[usenames]{color}
\usepackage{amsfonts,bm}
\usepackage{graphicx}

\newcommand{\be}{\begin{equation}}
\newcommand{\ee}{\end{equation}}
\newcommand{\bea}{\begin{eqnarray}\displaystyle}
\newcommand{\eea}{\end{eqnarray}}

\makeatletter
\@addtoreset{equation}{section}
\makeatother
\renewcommand{\theequation}{\thesection.\arabic{equation}}
\def\one{{\hbox{ 1\kern-.8mm l}}}
\def\zero{{\hbox{ 0\kern-1.5mm 0}}}

\begin{document}
\makeatletter
\@addtoreset{equation}{section}
\makeatother
\renewcommand{\theequation}{\thesection.\arabic{equation}}

\rightline{WITS-CTP-152}
\vspace{1.8truecm}

\vspace{15pt}


{\LARGE
\centerline{\bf  Topological String Correlators}
\centerline{\bf  from Matrix Models}
}  

\vskip.5cm 

 \thispagestyle{empty} \centerline{
     Robert de Mello Koch\footnote{{\tt robert@neo.phys.wits.ac.za},}
    and Lwazi Nkumane\footnote{{\tt lwazi.nkumane@gmail.com}}
                                                       }

\vspace{.4cm}
\centerline{{\it National Institute for Theoretical Physics ,}}
\centerline{{\it School of Physics and Centre for Theoretical Physics }}
\centerline{{\it University of Witwatersrand, Wits, 2050, } }
\centerline{{\it South Africa } }

\vspace{1.4truecm}

\thispagestyle{empty}

\centerline{\bf ABSTRACT}

We discuss how to compute connected matrix model correlators for operators related to the gravitational descendants of 
the puncture operator, for the topological A model on ${\bf P}^1$.
The relevant correlators are determined by recursion relations that follow from a systematic $1/N$ expansion of well chosen
Schwinger-Dyson equations.
Our results provide further compelling evidence for Gopakumar's proposed ``simplest gauge string duality'' between the
Gaussian matrix model and the topological A model on ${\bf P}^1$.

\vskip.4cm

\setcounter{page}{0}
\setcounter{tocdepth}{2}

\newpage

\setcounter{footnote}{0}

\linespread{1.1}
\parskip 4pt

\section{Introduction}\label{intro}

Gauge-string duality\cite{Maldacena:1997re,Gubser:1998bc,Witten:1998qj} represents a remarkable and deep connection 
between large $N$ gauge theories and theories of quantum gravity.
It would be nice to have simple tractable toy models of the correspondence that may shed light on the workings of the duality.
A natural candidate is to look for examples that relate topological gravity theories and large $N$ matrix models.   

With this motivation in mind, a number of connections between the combinatorics of Wick contractions in matrix
models and topological theories have been
uncovered\cite{Koch:2010zza,deMelloKoch:2011uq,deMelloKoch:2012ck,deMelloKoch:2012tb,Pasukonis:2013ts,Garner:2013qda,Geloun:2013kta,Koch:2014nka}.
This particular approach to the interpretation of the matrix models correlators, as well as certain counting problems,
is very reminiscent of the string interpretation of 
2d-Yang-Mills theory\cite{Gross:1992tu,Gross:1993hu,Gross:1993yt,Kimura:2008gs,Cordes:1994sd}.
In particular, the worldsheets that contribute to the correlators in the Gaussian matrix model are covering maps with
only three branch points and with simple ramification over one of the branch points\cite{Koch:2010zza}.
These are called ``clean'' Belyi maps.
This connection has been significantly extended by Gopakumar\cite{Gopakumar:2011ev} who has 
conjectured that the Gaussian matrix model is dual to the topological A-model on ${\bf P}^1$.
The duality between this Gaussian matrix model and the topological string certainly appears to deserve the title
of the simplest gauge string duality.
One may be skeptical about what can be learned from such a simple example of the duality.
The arguments of \cite{Gopakumar:2011ev} suggest that this correspondence is a concrete example
of closing the holes in ribbon graphs to obtain closed string worldsheets.
This is a concrete realization of the general approach to gauge string duality outlined 
in \cite{Gopakumar:2003ns,Gopakumar:2004qb,Gopakumar:2005fx}, and suitably modified to the Gaussian matrix model 
in \cite{Razamat:2008zr,Razamat:2009mc}.
These arguments convinced us that this example warrants further study.
In particular, we will further explore the relation between matrix model correlators and correlators in the topological string theory,
extending earlier works. 
For a review of the connection between topological string theories and matrix models, see \cite{Marino:2004eq}.
For background on the topological A-model on ${\bf P}^1$, see \cite{Okounkov:2002cz}.

A different but closely related matrix model dual to the topological A-model on ${\bf P}^1$ has been given by Eguchi and 
Yang\cite{Eguchi:1995er,Eguchi:1994in,Eguchi:1996tg}.
In this study, we will also compute the relevant correlators in the Eguchi-Yang model.

Motivated by the proposed ``simplest gauge-string duality'' we study correlation functions of operators of the form
${\rm Tr}(M^n {\rm ln}M)$ in this article.
In the Eguchi-Yang model, these operators are dual to gravitational descendants of the puncture 
operator\cite{Eguchi:1995er,Eguchi:1994in,Eguchi:1996tg}. 
For the Gaussian matrix model, Gopakumar and Pius\cite{rajesh2} have suggested these operators are again 
the natural candidates for operators dual to gravitational descendents of the puncture.
We will explore this suggestion in this article by computing the relevant connected correlators, both in the topological string
framework and in the matrix model description.
Basically, we explain how to derive recursion relations that determine these correlators.
The matrix model recursion relations follow upon applying a systematic $1/N$ expansion to well chosen Schwinger-Dyson equations.
We explain this connection in section 2.
For the topological string we use the recursion relations already 
described in\cite{Eguchi:1995er,Eguchi:1994in,Eguchi:1996tg,rajesh2}.
In section 3 we compute matrix model correlators of the Eguchi-Yang model and compare them to topological string correlators.
We find a complete and exact agreement between the two sets of correlators which is a convincing test of our methods.
Section 4 considers the question of the operators dual to gravitational descendants of the puncture operator
in the Gaussian matrix model.
Again, we find a set of operators that correctly reproduce the set of three point correlators.
In general there is a mismatch, which parallels results for correlators of gravitational descendants of the K\"ahler class\cite{rajesh2}.
The mismatch of \cite{rajesh2} was interpreted in terms of contact terms in the topological string theory.
We find that this argument also explains the mismatch we find, thereby providing highly non-trivial evidence in favor
of the gauge string duality proposed in \cite{Gopakumar:2011ev}.
We discuss these results in Section 5. 
The Appendices include a discussion of the recursion relations for both the Eguchi-Yang and the Gaussian matrix models, a
computation of the relevant topological string correlation functions and the computation of some matrix model correlators
using orthogonal polynomials.

\section{Matrix Model Correlators}\label{MM}

To illustrate our argument, we study correlators in the Gaussian matrix model
\bea
   \langle O\rangle \equiv \int [dM]\, e^{-{N\over 2}{\rm Tr} M^2}\, O
   \label{GMM}
\eea
where $M$ is an $N\times N$ Hermitian matrix.
The correlators that participate in the duality are connected correlators of the form
\bea
&& \gamma (2n_1,2n_2,\cdots 2n_l,\overline{2n_{l+1}},\cdots ,\overline{2n_{l+q}})=\cr
&&  \left\langle {\rm Tr}(M^{2n_1})\cdots {\rm Tr}(M^{2n_l})
         {\rm Tr}(M^{2n_{l+1}}\ln M)\cdots {\rm Tr}(M^{2n_{l+q}}\ln M)\right\rangle_{\rm conn}
\label{gencorfun}
\eea
Our notation should be clear from the above example: barred indices are associated with an insertion of ${\rm ln}M$ in the trace.
The correlators $\gamma (2n_1,2n_2,\cdots ,2n_k)$ have been computed combinatorially in \cite{Tutte} 
and by using the method of orthogonal polynomials in \cite{rajesh2}.
The combinatorial approach derives a recursion relation, which is then solved to obtain the correlators.
A central observation we use is that the recursion relation of \cite{Tutte} can be recovered by using a systematic 
$1/N$ expansion of a suitable matrix model Schwinger-Dyson equation.
The advantage of this approach is that it is simple to generalize the Schwinger-Dyson equation to allow for correlators that
include ${\rm ln}M$ insertions.
In this way, we derive a recursion relation for the general correlators
$\gamma (2n_1,\cdots 2n_l,\overline{2n_{l+1}},\cdots ,\overline{2n_{l+q}})$ introduced above.
It seems to be a highly non-trivial task to recover this recursion relation using the methods of \cite{Tutte}.

\subsection{Recursion relations from Schwinger-Dyson equations}

Correlators in the Hermittian  matrix model admit an expansion in ${1\over N^2}$.
For a correlator of the form
\bea
   \langle {\rm Tr}(M^{2n})\rangle = c_0 N+ c_1 N^{-1}+\cdots+c_p N^{1-2p}+\cdots
\eea
it is well known that the coefficient $c_g$ is given by the sum of all genus $g$ ribbon graphs.
We will use the notation
\bea
   c_p N^{1-2p}\leftrightarrow \langle {\rm Tr}(M^{2n})\rangle_p
\eea
For a correlator of the form
\bea
   \langle{\rm Tr}(M^{2n_1}){\rm Tr}(M^{2n_2})\rangle = c_0 N^{2}+ c_1
                                                                 +\cdots+c_p N^{2-2p}+\cdots
\label{twopoint}
\eea
things are a little more involved. 
The coefficient $c_0$ is obtained from disconnected ribbon graphs, that have two connected components.
Each component is a planar contribution to either $\langle {\rm Tr}(M^{2n_1})\rangle$ or $\langle {\rm Tr}(M^{2n_2})\rangle$.
The coefficient $c_1$ receives two types of contributions: a connected planar graph, and disconnected ribbon graphs that
have two connected components.
One of the connected components has genus 1 while the other is a planar graph.
Clearly, we need to distinguish between correlators and the connected piece of a correlator.
We do this with a subscript: $\langle {\rm Tr}(M^{2n_1}) {\rm Tr}(M^{2n_2})\rangle_{{\rm conn},g}$
is the connected contribution to $\langle {\rm Tr}(M^{2n_1}) {\rm Tr}(M^{2n_2})\rangle$ of genus $g$.
Using this notation, the coefficient $c_g$ in (\ref{twopoint}) is given by
\bea
    c_g N^{2-2g}= \langle {\rm Tr}(M^{2n_1}) {\rm Tr}(M^{2n_2})\rangle_{{\rm conn},g-1}
               +\sum_{\bar{g}}\langle {\rm Tr}(M^{2n_1})\rangle_{g-\bar{g}} 
                         \langle{\rm Tr}(M^{2n_2})\rangle_{\bar{g}}
\eea

The basic quantity we are interested in is $\gamma (2n_1,2n_2,\cdots,2n_k)$, where
\bea
   \langle {\rm Tr}(M^{2n_1}){\rm Tr}(M^{2n_2})\cdots {\rm Tr}(M^{2n_k})\rangle_{\rm conn,0}
=N^{2-k}\gamma (2n_1,2n_2,\cdots,2n_k)
\label{defofg}
\eea
The elements involved in the derivation of the recursion relation for general $k$ are all already present for $k=2$.
Since the $k=2$ case has fewer details, we discuss the $k=2$ relation in complete detail and are content to
outline the general $k$ argument.

Our starting point is the Schwinger-Dyson equation
\bea
      0=\int [dM]{d\over dM_{ij}}\left( (M^{2n_1-1})_{ij}{\rm Tr}(M^{2n_2})e^{-{N\over 2}{\rm Tr}(M^2)}\right)
\eea
which implies that
\bea
   0=\langle\sum_{r=0}^{2n_1-2} {\rm Tr}(M^{r}){\rm Tr}(M^{2n_1-r-2}){\rm Tr}(M^{2n_2})\rangle
+2n_2\langle {\rm Tr}(M^{2n_1+2n_2-2})\rangle\cr
 -N\langle {\rm Tr}(M^{2n_1}){\rm Tr}(M^{2n_2})\rangle
 \label{kistow}
\eea
To obtain a recursion relation for $\gamma(2n_1,2n_2)$ we consider a systematic large $N$ expansion of the above 
Schwinger-Dyson equation.
The leading order gives
\bea
   0&=&\sum_{r=0}^{2n_1-2} \langle {\rm Tr}(M^{r})\rangle_0\langle {\rm Tr}(M^{2n_1-r-2})\rangle_0 
                                             \langle{\rm Tr}(M^{2n_2})\rangle_0
 -N\langle {\rm Tr}(M^{2n_1})\rangle_0\langle {\rm Tr}(M^{2n_2})\rangle_0\cr
&=&\left[ \sum_{r=0}^{2n_1-2} \langle {\rm Tr}(M^{r})\rangle_0\langle {\rm Tr}(M^{2n_1-r-2})\rangle_0                                              
 -N\langle {\rm Tr}(M^{2n_1})\rangle_0\right] \langle {\rm Tr}(M^{2n_2})\rangle_0
\label{LO}
\eea

To see that this leading order equation is indeed obeyed, consider
\bea
      0=\int [dM]{d\over dM_{ij}}\left( (M^{2n_1-1})_{ij}\, e^{-{N\over 2}{\rm Tr}(M^2)}\right)
\eea
which implies that
\bea
   0=\langle\sum_{r=0}^{2n_1-2} {\rm Tr}(M^{r}){\rm Tr}(M^{2n_1-r-2})\rangle
 -N\langle {\rm Tr}(M^{2n_1})\rangle
  \label{anotherSD}
\eea
The leading order of this equation is
\bea
\sum_{r=0}^{2n_1-2} \langle {\rm Tr}(M^{r})\rangle_0\langle {\rm Tr}(M^{2n_1-r-2})\rangle_0                                              
 -N\langle {\rm Tr}(M^{2n_1})\rangle_0=0
\eea
which implies that (\ref{LO}) is obeyed.
The first subleading order of (\ref{anotherSD}) is
\bea
0=\sum_{r=0}^{2n_1-2} \langle {\rm Tr}(M^{r})\rangle_1\langle {\rm Tr}(M^{2n_1-r-2})\rangle_0
+\sum_{r=0}^{2n_1-2} \langle {\rm Tr}(M^{r})\rangle_0\langle {\rm Tr}(M^{2n_1-r-2})\rangle_1          \cr                                    
+\sum_{r=0}^{2n_1-2} \langle {\rm Tr}(M^{r}) {\rm Tr}(M^{2n_1-r-2})\rangle_{\rm conn,0}
 -N\langle {\rm Tr}(M^{2n_1})\rangle_1
\label{SL}
\eea
This equation will be important below.

Now consider the next to leading order of (\ref{kistow}) which gives
\bea
   0&&=
   \sum_{r=0}^{2n_1-2} \langle {\rm Tr}(M^{r})\rangle_1
                                        \langle {\rm Tr}(M^{2n_1-r-2})\rangle_0
                                              \langle {\rm Tr}(M^{2n_2})\rangle_0
+\sum_{r=0}^{2n_1-2} \langle{\rm Tr}(M^{r})\rangle_0
\langle {\rm Tr}(M^{2n_1-r-2})\rangle_1 \langle {\rm Tr}(M^{2n_2})\rangle_0\cr
&&+\sum_{r=0}^{2n_1-2} \langle{\rm Tr}(M^{r})\rangle_0
\langle {\rm Tr}(M^{2n_1-r-2})\rangle_0 \langle {\rm Tr}(M^{2n_2})\rangle_1
+\sum_{r=0}^{2n_1-2} \langle{\rm Tr}(M^{r})
{\rm Tr} (M^{2n_1-r-2})\rangle_{\rm conn,0}\langle {\rm Tr}(M^{2n_2})\rangle_0\cr
&&+\sum_{r=0}^{2n_1-2} \langle{\rm Tr}(M^{r})\rangle_0
\langle {\rm Tr}(M^{2n_1-r-2}) {\rm Tr}(M^{2n_2})\rangle_{\rm conn,0}
+\sum_{r=0}^{2n_1-2} \langle {\rm Tr} (M^{2n_1-r-2})\rangle_0
\langle{\rm Tr}(M^{r}) {\rm Tr}(M^{2n_2})\rangle_{\rm conn,0}\cr
&&+2n_2\langle {\rm Tr}(M^{2n_1+2n_2-2})\rangle_0
-N\langle {\rm Tr}(M^{2n_1}){\rm Tr}(M^{2n_2})\rangle_{\rm conn,0}
-N\langle {\rm Tr}(M^{2n_1})\rangle_1\langle {\rm Tr}(M^{2n_2})\rangle_0\cr
&&-N\langle {\rm Tr}(M^{2n_1})\rangle_0\langle {\rm Tr}(M^{2n_2})\rangle_1
\eea
The first, second, fourth and ninth terms above sum to zero thanks to (\ref{SL}).
This leaves
\bea
   0&&=
\sum_{r=0}^{2n_1-2} \langle{\rm Tr}(M^{r})\rangle_0
\langle {\rm Tr}(M^{2n_1-r-2})\rangle_0 \langle {\rm Tr}(M^{2n_2})\rangle_1
+\sum_{r=0}^{2n_1-2} \langle{\rm Tr}(M^{r})\rangle_0
\langle {\rm Tr}(M^{2n_1-r-2}) {\rm Tr}(M^{2n_2})\rangle_{\rm conn,0}\cr
&&+\sum_{r=0}^{2n_1-2} \langle {\rm Tr} (M^{2n_1-r-2})\rangle_0
\langle{\rm Tr}(M^{r}) {\rm Tr}(M^{2n_2})\rangle_{\rm conn,0}
+2n_2\langle {\rm Tr}(M^{2n_1+2n_2-2})\rangle_0\cr
&&-N\langle {\rm Tr}(M^{2n_1}){\rm Tr}(M^{2n_2})\rangle_{\rm conn,0}
-N\langle {\rm Tr}(M^{2n_1})\rangle_0\langle {\rm Tr}(M^{2n_2})\rangle_1
\eea
The first and last terms above sum to zero thanks to (\ref{LO}).
Consequently, we now obtain
\bea
   0&&=
\sum_{r=0}^{2n_1-2} \langle{\rm Tr}(M^{r})\rangle_0
\langle {\rm Tr}(M^{2n_1-r-2}) {\rm Tr}(M^{2n_2})\rangle_{\rm conn,0}
+\sum_{r=0}^{2n_1-2} \langle {\rm Tr} (M^{2n_1-r-2})\rangle_0
\langle{\rm Tr}(M^{r}) {\rm Tr}(M^{2n_2})\rangle_{\rm conn,0}\cr
&&+2n_2\langle {\rm Tr}(M^{2n_1+2n_2-2})\rangle_0
-N\langle {\rm Tr}(M^{2n_1}){\rm Tr}(M^{2n_2})\rangle_{\rm conn,0}
\eea
We can rewrite this as a recursion relation for $\gamma (2n_1,2n_2)$\footnote{In the first two terms it is clear that $r$
must be even to get a non-zero correlator.}
\bea
   0=
\sum_{r=0}^{n_1-1} \gamma(2r)\gamma(2n_1-2r-2,2n_2)
+\sum_{r=0}^{n_1-1} \gamma(2n_1-2r-2)\gamma(2r,2n_2)\cr
+2n_2\gamma(2n_1+2n_2-2)-\gamma(2n_1,2n_2)
\label{smplrecur}
\eea
This is in perfect agreement with the recursion relation obtained in \cite{Tutte}.

To obtain the general recursion relation, we start from
\bea
      0=\int [dM]{d\over dM_{ij}}\left( (M^{2n_1-1})_{ij}\prod_{j=2}^k{\rm Tr}(M^{2n_j})e^{-{N\over 2}{\rm Tr}(M^2)}\right)
\label{genrel}
\eea
and expand to the $k-1$th subleading order.
After freely make use of Schwinger-Dyson equations of the above form, that have fewer than $k-1$ traces, we recover the
general recursion relation of \cite{Tutte}.

\subsection{Including ${\rm ln}M$ insertions}

In this section we demonstrate that the logic of the previous subsection can be used to obtain a recursion relation for 
the correlators $\gamma (n_1,n_2,\cdots n_l,\bar{n}_{l+1},\cdots ,\bar{n}_{l+q})$.
It is again simplest to consider $k=2$ which illustrates all the features of the general case.
To obtain a recursion relation which determines $\gamma(n_1,\bar{n}_2)$ we start from the Schwinger-Dyson equation
\bea
      0=\int [dM]{d\over dM_{ij}}\left( (M^{2n_1-1})_{ij}{\rm Tr}(M^{2n_2}{\rm ln} M)e^{-{N\over 2}{\rm Tr}(M^2)}\right)
\eea
which implies that
\bea
   0=\langle\sum_{r=0}^{2n_1-2} {\rm Tr}(M^{r}){\rm Tr}(M^{2n_1-r-2}){\rm Tr}(M^{2n_2}{\rm ln} M)\rangle
+2n_2\langle {\rm Tr}(M^{2n_1+2n_2-2}{\rm ln} M)\rangle\cr
+\langle {\rm Tr}(M^{2n_1+2n_2-2})\rangle
 -N\langle {\rm Tr}(M^{2n_1}){\rm Tr}(M^{2n_2}{\rm ln} M)\rangle
\eea
Expanding this Schwinger-Dyson equation in a ${1\over N}$ expansion and making use of (\ref{anotherSD}) in much the same way
that we did above, it is straight forward to obtain
\bea
\gamma (2n_1,\overline{2n_2})=\sum_{r=0}^{n_1-1}\gamma (2n_1-2r-2,\overline{2n_2})\gamma (2r)
+\sum_{r=0}^{n_1-1}\gamma (2r,\overline{2n_2})\gamma (2n_1-2r-2)\cr
+2n_2\gamma(\overline{2n_1+2n_2-2})+\gamma(2n_1+2n_2-2)
\eea
This looks very similar to the recursion (\ref{smplrecur}).
The first two ``trace splitting'' terms on the right hand side are an exact match.
The only difference is in the joining term - there is a term where the two traces have joined with the log in a single
trace (the term with coefficient $2n_2$) as well as a term in which the log has been absorbed in the joining (the term with
coefficient 1).
This relation is rather general.
For example, at $k=3$ the recursion from \cite{Tutte} reads
\bea
   0=
\sum_{r=0}^{n_1-1} \gamma(2r)\gamma(2n_1-2r-2,2n_2,2n_3)+\sum_{r=0}^{n_1-1} \gamma(2r,2n_3)
\gamma(2n_1-2r-2,2n_2)\cr
+\sum_{r=0}^{n_1-1} \gamma(2n_1-2r-2,2n_3)\gamma(2r,2n_2)+\sum_{r=0}^{n_1-1} \gamma(2n_1-2r-2)
\gamma(2r,2n_2,2n_3)\cr
+2n_2\gamma(2n_1+2n_2-2,2n_3)+2n_3\gamma(2n_1+2n_3-2,2n_2)-\gamma(2n_1,2n_2,2n_3)
\eea
One of the recursion relations\footnote{The recursion given can be used to determine 
$\gamma(2n_1,\overline{2 n_2},\overline{2 n_3})$. There are two other independent recursions which would determine
$\gamma(\overline{2 n_1},\overline{2 n_2},\overline{2 n_3})$ and
$\gamma(2n_1,2n_2,\overline{2 n_3})$.} relevant for the log insertions is
\bea
   0=
\sum_{r=0}^{n_1-1} \gamma(2r)\gamma(2n_1-2r-2,\overline{2 n_2},\overline{2 n_3})
+\sum_{r=0}^{n_1-1} \gamma(2r,\overline{2 n_3})\gamma(2n_1-2r-2,\overline{2 n_2})\cr
+\sum_{r=0}^{n_1-1} \gamma(2n_1-2r-2,\overline{2 n_3})\gamma(2r,\overline{2 n_2})
+\sum_{r=0}^{n_1-1} \gamma(2n_1-2r-2)\gamma(2r,\overline{2 n_2},\overline{2 n_3})\cr
+2n_2\gamma(\overline{2n_1+2n_2-2},2\bar n_3)+\gamma(2n_1+2n_2-2,\overline{2n_3})\cr
+2n_3\gamma(\overline{2n_1+2n_3-2},\overline{2n_2})
+\gamma(2n_1+2n_3-2,\overline{2n_2})
-\gamma(2n_1,\overline{2n_2},\overline{2 n_3})
\eea
The relation between these two recursions is exactly as we described it above.

\subsection{Solution to the recursion relation}

The solution to the recursion relation for $\gamma (2n_1,2n_2,\cdots,2n_k)$ has already been obtained in \cite{Tutte}.
The result is
\bea
   \gamma (2n_1,2n_2,\cdots,2n_k)\equiv N^{k-2}\langle\prod_{i=1}^k {\rm Tr}(M^{2n_i})\rangle_{\rm conn,0}
              ={(n-1)!\over (n-k+2)!}\prod_{i=1}^k {(2n_i)!\over n_i! (n_i-1)!}
\eea
where $n=\sum_{i=1}^k n_i$.
The solution of the general recursion relation is now straightforward.
Indeed, we have verified that the formula for the 
$\gamma (2n_1,\cdots 2n_l,\overline{2n_{l+1}},\cdots ,\overline{2n_{l+q}})$,
obtained through analytic continuation of $\gamma (2n_1,\cdots 2n_l,2n_{l+1},\cdots ,2n_{l+q})$, solves the
relevant recursion relation.
As an example,
\bea
   \gamma (2n_1,2n_2,&\cdots&,2n_{k-1},\overline{2 n_k})
\equiv N^{k-2}\langle\prod_{i=1}^{k-1} {\rm Tr}(M^{2n_i}){\rm Tr}(M^{2n_k}{\rm ln}M)\rangle_{\rm conn,0}\cr
&=&\lim_{\epsilon\to 0}{N^{k-2}\over 2\epsilon}
\langle\prod_{i=1}^{k-1} {\rm Tr}(M^{2n_i}){\rm Tr}(M^{2n_k+2\epsilon})\rangle_{\rm conn,0}\cr
&=&{1\over 2}{(n-1)!\over (n-k+2)!}\prod_{i=1}^k {(2n_i)!\over n_i! (n_i-1)!}\times\cr
&&\,\,\,\,\,\times
\left(
\sum_{j=1}^{n-1}{1\over j}-\sum_{j=1}^{n-k+2}{1\over j}
+2\sum_{j=1}^{2 n_k}{1\over j}
-\sum_{j=1}^{n_k}{1\over j}-\sum_{j=1}^{n_k-1}{1\over j}
\right)
\eea

\section{Topological String Correlators for the A Model on ${\bf P}^1$}\label{TS}

Eguchi and Yang proposed a matrix model which reproduces the A-model on 
${\bf P}^1$\cite{Eguchi:1995er,Eguchi:1994in,Eguchi:1996tg}.
The observables of the theory are the puncture operator $P$ and the K\"ahler class $Q$,
as well as their gravitational descendents $\sigma_n (P)$ and $\sigma_n (Q)$.
The action of the matrix model 
\bea
   Z=\int [dM] e^{N{\rm Tr} V(M)}
\eea
is 
\bea
   {\rm Tr}V(M) = -2 {\rm Tr} (M{\rm ln}M -M)+\sum_{n=1} 2t_{n,P}{\rm Tr}(M^n {\rm ln}M-c_n M^n)
                          +\sum_{n=1}t_{n-1,Q}{{\rm Tr} M^n\over n}
\label{EYMM}
\eea
where
\bea
   c_n=\sum_{j=1}^n {1\over j}
\eea
The correlators of the topological string are determined by solving recursion relations.
If one puts all the coupling to zero, one finds\cite{Eguchi:1995er}\footnote{$\langle\cdots\rangle_g$ denotes the genus $g$
contribution to the topological string correlator. $\langle\cdots\rangle_{g,d}$ denotes the genus $g$ contribution to the
topological string correlator coming from degree $d$ maps.}
\bea
   \langle\sigma_{2m}(Q)\rangle_0 ={(2m)!\over (m+1)!(m+1)!}
   \qquad
  \langle\sigma_{2m+1}(Q)\rangle_0 =0
 \label{ts1}
\eea
\bea
   \langle\sigma_{2m+1}(P)\rangle_0 =-2c_{m+1}{(2m+1)!\over (m+1)!(m+1)!}
   \qquad
   \langle\sigma_{2m}(P)\rangle_0 =0
   \label{ts2}
\eea
As a consequence of ghost number conservation, the above correlators only get a contribution from maps of degree $m+1$.
Using these it is now straight forward to find\cite{rajesh2}
\bea
\langle\sigma_{2m_1}(Q)\sigma_{2m_2}(Q)\rangle_0
={1\over m_1+m_2+1}{(2m_1)!\over m_1!m_1!}{(2m_2)!\over m_2!m_2!}  \label{ts3}
\eea
\bea
\langle\sigma_{2m_1-1}(Q)\sigma_{2m_2-1}(Q)\rangle_0 
={1\over 4( m_1+m_2)}{(2m_1)!\over m_1!m_1!}{(2m_2)!\over m_2!m_2!}  \label{ts4}
\eea
\bea
\langle\sigma_{2m_1-1}(Q)\sigma_{2m_2}(Q)\rangle_0 
=0
\eea
as well as (see the Appendix)
\bea
   \langle\sigma_{2m_1}(Q)\sigma_{2m_2+1}(P)\rangle_0 
        = {1\over m_1+m_2+1}{(2m_1)!\over m_1! m_1!}{(2m_2+1)!\over m_2! m_2!}
         \left[-2c_{m_2}-{1\over m_1+m_2+1}\right]  \label{ts5}
\eea
\bea
   \langle\sigma_{2m_1+1}(Q)\sigma_{2m_2}(P)\rangle_0 
        = {1\over 2(m_1+m_2+1)}{(2m_1+2)!\over (m_1+1)! (m_1+1)!}{(2m_2)!\over m_2! (m_2-1)!}\times\cr
        \times \left[-2c_{m_2}+{1\over m_2}-{1\over m_1+m_2+1}\right] 
\eea
Using the identifications
\bea
   \sigma_{m}(Q)\leftrightarrow  {{\rm Tr} M^{m+1}\over m+1}
\qquad
   \sigma_{m}(P)\leftrightarrow  2{\rm Tr} \left( M^{m}{\rm ln} M-c_{m}M^{m}\right)
\label{EYHmap}
\eea
we can use the topological string correlators (\ref{ts1}) - (\ref{ts5}) to solve for the expected matrix model correlators.
%
%
In this way, the topological string theory makes the following predictions (all $\gamma$s predicted using the topological
string correlators are hatted)
\bea
  \hat\gamma(2m+1) ={(2m+1)!\over (m+1)!(m+1)!}\qquad
  \hat\gamma(2m)=0\label{mmc1}
\eea
\bea
\hat\gamma(\overline{2m+1}) =(c_{2m+1}-c_{m+1}){(2m+1)!\over (m+1)!(m+1)!}\qquad
\hat\gamma(\overline{2m})=0
\label{mmc2}
\eea
\bea
\hat\gamma(2m+1,2n) =0\label{mmc3}
\eea
\bea
\hat\gamma(2m+1,2n+1)=
{1\over m+n+1}{(2m+1)!\over m! m!}{(2n+1)!\over n! n!}\label{mmc4}
\eea
\bea
\hat\gamma(2m,2n)=
{mn\over m+n}{(2m)!\over m! m!}{(2n)!\over n! n!}\label{mmc5}
\eea
\bea
\hat\gamma(2n+1,\overline{2m+1})
=\left({c_{2m_2+1}-c_{m_2}\over m_1+m_2+1}-{1\over 2(m_1+m_2+1)^2}\right){(2m_1+1)!\over m_1!m_1!}
          {(2m_2+1)!\over m_2!m_2!}\label{mmc6}
\eea
Using the methods of section 2, we will derive recursion relations in this section that will test these predictions.
Starting from
\bea
0=\int [dM] {d\over dM_{ij}}\left( (M^n)_{ij}e^{N{\rm Tr}V}\right)
\eea
we find (recall that we have set all couplings to zero)
\bea
2N\langle {\rm Tr}(M^n{\rm ln} M)\rangle =\sum_{r=0}^{n-1}\langle {\rm Tr}(M^{n-1-r}){\rm Tr}(M^{r})\rangle
\label{leadingeguchi}
\eea
The leading order of this Schwinger-Dyson equation implies that
\bea
2\gamma (\bar n)=\sum_{r=0}^{n-1}\gamma (r)\gamma (n-1-r)\label{eguchi1}
\eea
Inserting (\ref{mmc1}) into (\ref{eguchi1}) we recover (\ref{mmc2}).
Now consider the two point correlators.
Starting from
\bea
0=\int [dM] {d\over dM_{ij}}\left( (M^n)_{ij} {\rm Tr}(M^m)e^{N{\rm Tr}V}\right)
\eea
we find
\bea
2N\langle {\rm Tr}(M^m) {\rm Tr}(M^n{\rm ln} M)\rangle 
=\sum_{r=0}^{n-1}\langle {\rm Tr}(M^{n-1-r}){\rm Tr}(M^{r}){\rm Tr}(M^m)\rangle
+m\langle {\rm Tr}(M^{n+m-1})\rangle
\eea
Expanding this Schwinger-Dyson equation to subleading order, after using both the leading and subleading order of
(\ref{leadingeguchi}), we find
\bea
  2\gamma (m,\bar n)=\sum_{r=0}^{n-1}\left(\gamma (r,m)\gamma (n-1-r)+\gamma (n-1-r,m)\gamma(r)\right)
+m\gamma (n+m-1)
\label{bigrecur}
\eea
Using (\ref{mmc3})-(\ref{mmc5}) in (\ref{bigrecur}) we recover (\ref{mmc6}).
For all of the examples we have considered, the matrix model correlators and topological string theory correlators 
are in perfect agreement.
Notice that there is a natural way to understand the result (\ref{mmc6}) by analytic continuation.
Indeed, it is simple to verify that
\bea
\gamma(n,\overline{m})&&=\langle {\rm Tr}(M^{n}) {\rm Tr}( M^{m}{\rm ln} M)\rangle_0\cr
&&={d\over d\epsilon}\langle {\rm Tr}(M^{n}) {\rm Tr}( M^{m+\epsilon})\rangle_0\Big|_{\epsilon=0}\cr
&&={d\over d\epsilon}\gamma(n,m+\epsilon)\Big|_{\epsilon=0}\label{continue}
\eea
In general we expect the $\gamma(n_1,n_2,\cdots,n_{k-1},n_k)$ and $\gamma(n_1,n_2,\cdots,n_{k-1},\bar n_k)$
that solve the recursion relations, to be related by analytic continuation.
With the assumption (\ref{continue}), we have managed to solve the Schwinger-Dyson equations for the 
$\gamma (n_1,n_2,\cdots,n_k)$ and have verified that the correlators of the gravitational descendants 
computed in the matrix model agree with the same correlators computed using the topological string.
The translation between the two is given by the first of (\ref{EYHmap}).
As an example, the topological string correlators\cite{rajesh2}
\bea
   \langle \prod_{i=1}^k\sigma_{2m_i}(Q)\rangle_0 = (d+1)^{k-3}\prod_{i=1}^k {(2 m_i)!\over m_i! m_i!}\qquad
   d=\sum_{i=1}^k m_i\label{genTS}
\eea
predict
\bea
   \gamma (2m_1+1,2m_2+1,\cdots ,2m_k+1) = (d+1)^{k-3}\prod_{i=1}^k {(2 m_i+1)!\over m_i! m_i!}
\label{genans}
\eea
Setting
\bea
   \gamma (2m_1+1,\cdots ,\overline{2m_k+1}) = {d\over d\epsilon}   
   \gamma (2m_1+1,\cdots ,2m_k+1+\epsilon)\Big|_{\epsilon=0}
\eea
we find that $\gamma (2m_1+1,2m_2+1,\cdots ,2m_k+1)$ and $\gamma (2m_1+1,\cdots ,\overline{2m_k+1})$ do indeed
satisfy the Schwinger-Dyson equations.
Thus, the Eguchi-Yang matrix model does indeed correctly reproduce the correlators of the topological string theory.
Apart from this agreement, we now have good evidence that the recursion relations we have obtained for the correlators that
include ${\rm Tr} M^n{\rm ln}M$ insertions are indeed correct.

\section{The Simplest Gauge String Duality}\label{dual}

In this section we return to the Gaussian matrix model. 
The combinatorics of the Wick contractions in the Gaussian matrix model suggests that correlators are computed by a
sum over branched covers from a genus $g$ worldsheet to a target ${\bf P}^1$ ``spacetime''.
These holomophic maps have exactly three branchpoints \cite{Koch:2010zza}, and are known as Belyi maps.
Using mathematical results from \cite{Mulase}, the explicit form of these Belyi maps has been given in \cite{Gopakumar:2011ev}.
From the detailed form of the map, it is clear that 
$\langle \prod_{i=1}^{n}{\rm Tr}(M^{2k_i})\rangle_{\rm conn,0}$ only receives 
planar contributions only from degree $d=\sum_i k_i$ maps.
As a consequence of the ghost number conservation law (see \cite{Eguchi:1996tg} for more details) the genus
0 contribution to the correlator of descendants of the K\"ahler class
\bea
   \langle\sigma_{2k_1-1}(Q)\sigma_{2k_2-1}(Q)\sigma_{2k_3}(Q)\cdots \sigma_{2k_n}(Q)\rangle
\eea
only receives contributions from maps of degree
\bea
   d=\sum_{i=1}^k k_i
\eea
which strongly suggests \cite{Gopakumar:2011ev,rajesh2} the rough identification
\bea
\sigma_{2k}(Q)\sim {\rm Tr} \left( M^{2k}\right)
\eea
With this identification, the two point and three point functions in the Gaussian matrix model 
and in the topological string theory match\cite{Gopakumar:2011ev} for any $k_i$
\bea
\langle {1\over 2k_1}{\rm Tr} M^{2k_1}{1\over 2k_2}{\rm Tr} M^{2k_2}\rangle_{\rm conn,0}
={1\over 4(k_1+k_2)}{(2k_1)!\over (k_1!)^2}{(2k_2)!\over (k_2!)^2}\cr
=\langle \sigma_{2k_1-1}(Q)\sigma_{2k_2-1}(Q)\rangle_0
\eea
\bea
\langle {1\over 2k_1+1}{\rm Tr} M^{2k_1+1}{1\over 2k_2+1}{\rm Tr} M^{2k_2+1}\rangle_{\rm conn,0}
={1\over k_1+k_2+1}{(2k_1)!\over (k_1!)^2}{(2k_2)!\over (k_2!)^2}\cr
=\langle \sigma_{2k_1}(Q)\sigma_{2k_2}(Q)\rangle_0
\eea
\bea
\langle {1\over 2k_1}{\rm Tr} M^{2k_1}{1\over 2k_2}{\rm Tr} M^{2k_2}{1\over k_3}{\rm Tr} M^{2k_3}\rangle_{\rm conn,0}
={1\over 4}{(2k_1)!\over k_1!k_1!}{(2k_2)!\over k_2!k_2!}{(2k_3)!\over k_3!k_3!}\cr
=\langle\sigma_{2k_1-1}(Q)\sigma_{2k_2-1}(Q)\sigma_{2k_3}(Q)\rangle_0
\eea
\bea
\langle {1\over 2k_1+1}{\rm Tr} M^{2k_1+1}{1\over 2k_2+1}{\rm Tr} M^{2k_2+1}{1\over k_3}
{\rm Tr} M^{2k_3}\rangle_{\rm conn,0}
={(2k_1)!\over k_1!k_1!}{(2k_2)!\over k_2!k_2!}{(2k_3)!\over k_3!k_3!}\cr
=\langle\sigma_{2k_1}(Q)\sigma_{2k_2}(Q)\sigma_{2k_3}(Q)\rangle_0
\eea
The general map is as follows
\bea
\langle {{\rm Tr} M^{2k_1}\over 2k_1}{{\rm Tr} M^{2k_2}\over 2k_2}{{\rm Tr} M^{2k_3}\over k_3}\cdots 
            {{\rm Tr} M^{2k_n}\over  k_n}\rangle_{\rm conn,0}\,\,\leftrightarrow\,\,
\langle\sigma_{2k_1-1}(Q)\sigma_{2k_2-1}(Q)\sigma_{2k_3}\cdots \sigma_{2k_n}(Q)\rangle_0
\eea
These correlators appear to agree, up to contact terms\cite{rajesh2}.
Notice that when computing the topological string correlator, we treat two of the vertex operators differently to the remaining
operators.
The need for this is simply to ensure that the degree of the Belyi maps contributing to the matrix model correlator matches the
degree of the holomorphic map contributing to the topological string theory correlator - so that the closed surface obtained
after closing the holes in the ribbon graph is the topological closed string worldsheet.
This is very much like usual worldsheet descriptions, where we fix the positions of three vertex operators.
Here a difference of 2 in the degree of the two maps is being accounted for by putting two of the vertex 
operators at fixed points on the worldsheet.
These results present a rather compelling case for the duality.

We would now like to propose the operators dual to the gravitational descendents of the puncture.
In \cite{rajesh2} it was suggested that $\sigma_n(P)\sim {\rm Tr}(M^n {\rm ln} M)$, which we will see, is essentially correct.
We are again guided by matching correlators.
It is the three point correlators that do not receive contact term contributions\cite{rajesh2}, so that we should require that the
three point topological string correlators match with the corresponding matrix model correlators.

To reproduce the complete set of three point correlators, we propose that
\bea
\sigma_{2k_3+1}(P)\leftrightarrow {2k_3+1\over k_3}\left( {\rm Tr} \left( 2M^{2k_3}{\rm ln} M
             -\left[{1\over k_3}+2c_{2k_3}\right]M^{2k_3}\right)\right)\label{gravdes}
\eea
Notice that there is a shift of 1 between the level of the gravitational descendant ($2k_3+1$) of the puncture and the power of 
the matrix ($2k_3$) in the matrix model operator.
To see why this shift is necessary, recall that when all the couplings are set to zero the genus $g$ correlation functions 
$\langle \sigma_{n_1}(O_{\alpha_1})\cdots \sigma_{n_s}(O_{\alpha_s})\rangle_g$
only receive contributions from holomorphic maps of degree $d$ with
\bea
   2d + 2(g-1)=\sum_{i=1}^s (n_i+q_{\alpha_i}-1)
\eea
Since the $U(1)$ charge of the puncture $q_P=0$, including $\sigma_{2k_3+1}(P)$ in the correlator adds $k_3$ to the degree.
This matches the matrix model computation since including ${\rm Tr}(M^{2k_3})$ (and hence also ${\rm Tr}(M^{2k_3}{\rm ln}M)$)
adds degree $k_3$ to the Belyi maps summed to reproduce the matrix model correlator.
It is now straight forward to verify that
\bea
&&\langle \sigma_{2k_1-1}(Q)\sigma_{2k_2-1}(Q)\sigma_{2k_3 +1}(P)\rangle =\cr
&&\langle
{{\rm Tr}M^{2k_1}\over 2k_1}{{\rm Tr}M^{2k_2}\over 2k_2}
{2k_3+1\over k_3}\left( {\rm Tr} \left( 2M^{2k_3}{\rm ln} M
             -\left[{1\over k_3}+2c_{2k_3}\right]M^{2k_3}\right)\right)\rangle_{\rm conn,0}
\eea
The fact that (\ref{gravdes}) reproduces an infinite number of correlators is strong evidence that it is indeed on the right
track. 

Lets now consider more general correlators. 
Comparing the topological string correlator
\bea
\langle \sigma_{2k_1-1}(Q)\sigma_{2k_2-1}(Q)\prod_{i=3}^n \sigma_{2k_i}(Q)\sigma_{2k+1}(P)\rangle \cr
={d^{n-2}\over 4}\prod_{i=1}^n {(2k_i)!\over k_i!k_i!}{(2k+1)!\over k!k!}
\left(-2c_{k}+{n-2\over d}\right)
\eea
to the matrix model correlator
\bea
&& 
\left\langle{1\over 4}\prod_{i=1}^n {{\rm Tr}M^{2k_i}\over k_i}
{2k+1\over k} {\rm Tr} \left( 2M^{2k}{\rm ln} M
             -\left[{1\over k}+2c_{2k}\right]M^{2k}\right)\right\rangle_{\rm conn,0}\cr
&&={(d-1)!\over (d-n+1)!}{1\over 4}\prod_{i=1}^n {(2k_i)!\over k_i!k_i!}
       {(2k+1)!\over k!k!}\Bigg(-2c_{k}+c_{d-1}-c_{d-n+1}\Bigg)
\eea
where $d=k_1+\cdots+k_n+k$, we see two sources of mismatch.
First, there is an overall factor of $d^{n-2}$ versus ${(d-1)!\over (d-n+1)!}$.
In the large $d$ limit these two factors are identical.
This factor has an elegant interpretation in terms of contact term corrections, as explained in \cite{rajesh2}.
Further, in the large $d$ limit, we can replace
\bea
{n-2\over d} \quad \to \quad 0
\eea
in the topological string correlator and
\bea
  c_{d-1}-c_{d-n+1} \approx \int_1^{d-1}{dx\over x}-\int_1^{d-n+1}{dx\over x}={\rm ln}{d-1\over d-n+1}\to 0
\eea
in the matrix model correlator.
Both of these numbers grow linearly with the number of operators in the correlator that are gravitational descendents 
of the puncture operator.
Consequently, this term continues to go to zero in the large $d$ limit for correlators with an arbitrary but finite number
of  the gravitational descendants of the puncture.
This is a BMN like limit.
In this limit there is perfect agreement between the topological string correlators and the matrix model correlators.

\section{Discussion}\label{conclude}

Motivated by what may be the simplest proposal for a gauge string duality\cite{Gopakumar:2011ev},
we have explored the relation between correlators of the topological A-model on ${\bf P}^1$ and the correlators of the Gaussian
matrix model, as well as the correlators of the closely related model of Eguchi and Yang.
For the Gaussian matrix model, we have written down operators in the matrix model whose three point function matches
the three point function of gravitational descendants in the topological A-model, thereby adding to the dictionary between
observables of the Gaussian matrix model and the topological string.

To compute the required matrix model correlators, we have developed recursion relations that allow the computation of
correlators of the form (\ref{gencorfun}) in any matrix model.
As we have explained, these recursion relations follow from a systematic $1/N$ expansion of the Schwinger-Dyson equations.

Although three point functions computed in the matrix models and the topological string theory agree, there are
discrepancies between the two complete sets of correlation functions.
The disagreement between matrix model correlators and topological string correlators was already discovered and
discussed in \cite{rajesh2}.
Lets review this proposal briefly.
The source of the disagreement lies in contact terms: there is no operator product expansion in the matrix model that
describes what happens when two traces coincide.
To get agreement one needs to add the separate contributions from the fusing of two matrix operators by hand.
In terms of equations, it is simplest to use the alternative normalizations
\bea
   \tilde\sigma_{2k}(Q)={(k!)^2\over (2k)!}\sigma_{2k}(Q)\qquad O_{2k}^Q={1\over k}{(k!)^2\over (2k)!}{\rm Tr} (M^{2k})
\eea
We now find that the topological string correlators
\bea
\langle \tilde \sigma_{2k_1-1}(Q)\tilde \sigma_{2k_2-1}(Q)\prod_{i=3}^n \sigma_{2k_i}(Q)\rangle
={d^{n-3}\over 4}
\eea
and the matrix model correlators
\bea
\left\langle {O^Q_{2k_1}\over 2}{O^Q_{2k_2}\over 2}\prod_{i=3}^n O^Q_{2k_i}\right\rangle_{\rm conn,0}
={1\over 4}{(d-1)!\over (d-n+2)!}
\eea
are related as
\bea
   \langle \tilde \sigma_{2k_1-1}(Q)\tilde \sigma_{2k_2-1}(Q)\prod_{i=3}^n \tilde\sigma_{2k_i}(Q)\rangle =
  \sum_{m=3}^n \tilde S^{(m-2)}_{n-2}
\left\langle  {O^Q_{2k_1}\over 2}{O^Q_{2k_2}\over 2}\prod_{i=3}^m O^Q_{2k_i}\right\rangle_{\rm conn,0}
\eea
where the Stirling number of the second kind $\tilde S^{(m-2)}_{n-2}$ counts the number of ways to partition
$(3,4,\cdots,n)$ into $m-2$ sets.
The above equality follows from the identity
\bea
   d^{n-3}=\sum_{m=3}^n \tilde S_{n-2}^{(m-2)}{(d-1)!\over (d-m+2)!}\label{frstident}
\eea
This explanation of the disagreement is clearly not special to the gravitational descendants of the K\"ahler class
so one should also expect this argument to resolve any mismatches  for correlators involving the gravitational descendants 
of the puncture.
This is a highly non-trivial prediction that we can now test.
Using the puncture normalizations
\bea
   \tilde\sigma_{2k+1}(P)={(k!)^2\over (2k+1)!}\sigma_{2k+1}(P)\qquad
O^P_{2k+1}={k! (k-1)!\over (2k)!} {\rm Tr} \left( 2M^{2k}{\rm ln} M
             -\left[{1\over k}+2c_{2k}\right]M^{2k}\right)\nonumber
\eea
we find
\bea
\langle \tilde\sigma_{2k_1-1}(Q)\tilde\sigma_{2k_2-1}(Q)\prod_{i=3}^n \tilde\sigma_{2k_i}(Q)\tilde\sigma_{2k+1}(P)\rangle
={d^{n-2}\over 4}\left(-2c_{k}+{n-2\over d}\right)
\eea
and
\bea
\left\langle {1\over 4}\prod_{i=1}^n O_{2k_i}^Q O_{2k}^P \right\rangle_{\rm conn}
={1\over 4}{(d-1)!\over (d-n+1)!}\Bigg(-2c_{k}+c_{d-1}-c_{d-n+1}\Bigg)
\eea
These two correlators are related by
\bea
   \langle \tilde \sigma_{2k_1-1}(Q)\tilde \sigma_{2k_2-1}(Q)\prod_{i=3}^n \tilde\sigma_{2k_i}(Q)\tilde\sigma_{2k+1}(P)\rangle =
  \sum_{m=3}^{n+1} \tilde S^{(m-2)}_{n-1}\left\langle {1\over 4}\prod_{i=1}^m O^Q_{2k_i}O^P_{2k}\right\rangle_{\rm conn}
\label{contactforpuncture}
\eea
The above equality uses the identity
\bea
  (n-2)d^{n-3}=\sum_{m=3}^{n+1}\tilde S^{(m-2)}_{n-1}{(d-1)!\over (d-m+2)!}(c_{d-1}-c_{d-m+2})
\eea
which easily follows by differentiating (\ref{frstident}) with respect to $d$. 
The derivative of (\ref{frstident}) must also hold because (\ref{frstident}) is just an equality between two polynomials in $d$.
This demonstrates that after contact terms are accounted for as in (\ref{contactforpuncture}), there is exact agreement 
between the topological string and matrix model correlators, providing highly nontrivial evidence for the proposal 
of \cite{Gopakumar:2011ev,rajesh2}.

In \cite{rajesh2} it was already pointed out that the correlators of gravitational descendents of the Kahler class computed in 
the Gaussian matrix model and the topological string agree in the limit of large descendent level.
This is the regime in which the contributing ribbon graphs have a large number of edges and faces.
In this limit, contact term contributions are subleading.
Consequently, when the descendent levels are large we have perfect agreement between the Gaussian matrix model 
correlators and the topological string correlators, with no need for contact term corrections.
Thus, we are recovering a correct continuum description of the moduli space \cite{Razamat:2008zr}.

Our recursion relations can further be expanded in the $1/N$ expansion. 
This would allow the study of higher genus contributions and hence of the mixing of single and double traces in the connected
correlators.
Studies of these corelators would allow us to explore the correspondence between the topological A-model and the Gaussian 
matrix model beyond genus zero, which is clearly required to prove this example of gauge string duality.

{\vskip 0.2cm}

\noindent
{\it Acknowledgements:}
We would like to thank Sanjaye Ramgoolam, Joao Rodrigues and especially Rajesh Gopakumar for useful discussions and comments
on the manuscript.
This work is based upon research supported by the South African Research Chairs
Initiative of the Department of Science and Technology and National Research Foundation.
Any opinion, findings and conclusions or recommendations expressed in this material
are those of the authors and therefore the NRF and DST do not accept any liability
with regard thereto.

\begin{appendix}

\section{Recursion relations for connected matrix model correlators}\label{RR}

In this section we summarize some of the recursion relations obtained from a systematic ${1\over N}$ expansion of the
Schwinger-Dyson equations. 

\subsection{The Gaussian Model}

The corelators of this model are defined by (\ref{GMM}).
Starting from the Schwinger-Dyson equation (\ref{genrel}) and proceeding as described in section 2,
we obtain the following recursion relation
\bea
   \gamma(2n_1,2n_2,\cdots,2n_k)=
\sum_P\sum_{j=0}^{n_1-1}\gamma(2j,P)\gamma(2n_1-2j-2,\tilde P)\cr
+\sum_{r=2}^k 2n_r \gamma (2n_1+2n_r-2,S'_r)
\eea
In the above equation, $S$ stands for the set $\{ 2n_2,2n_3,\cdots, 2n_k\}$.
The first term on the right hand side is a ``splitting term'' in which we split the set 
${2n_1-2}\cup S=\{2n_1-2,2n_2,\cdots,2n_k\}$ into two subsets and use each to define a $\gamma$.
This corresponds to splitting the trace over $M^{2n_1}$ into two traces, so that the first
term on the right hand side comes from terms in the Schwinger-Dyson equation involving
$k+1$ traces.
We partition $S$ into $P$ and $\tilde P$, and split $2n_1\to 2j,2n_1-2-2j$; the two parts of
${2n_1-2}\cup S$ are $2j\cup P$ and $2n_1-2j-2\cup \tilde P$. The two sums run over all possible
ways of breaking $S$ into $P$ and $\tilde P$, and over all possible values of $j$.
The second term on the right hand side is a ``joining term'' - it comes from a term in the
Schwinger-Dyson equation involving $k-1$ traces.
The set $S'_r$ is given by removing $2n_r$ from $S$.

The recursion relation obtained when we allow logs in the traces is very similar.
There are basically two changes.
First, the set $S$ includes both integers and barred integers (the terms with log insertions).
Second there is a new trace splitting term.
As an example of a recursion relation including logs, consider
\bea
   \gamma(2n_1,2n_2,\cdots ,2n_{k-2},\overline{2n_{k-1}},\overline{2n_k})=
\sum_P\sum_{j=0}^{n_1-1}\gamma(2j,P)\gamma(2n_1-2j-2,\tilde P)\cr
+\sum_{r=2}^{k-2} 2n_r \gamma (2n_1+2n_r-2,S'_r)
+\sum_{r=k-1}^{k} 2n_r \gamma (\overline{2n_1+2n_r-2},S'_r)\cr
+\sum_{r=k-1}^{k} \gamma (2n_1+2n_r-2,S'_r)
\eea
In the last term on the right hand side of the above equation, one barred index has become unbarred.
The remaining terms all agree on the barred and unbarred indices.

\subsection{Eguchi-Yang matrix model}

The model is defined by the potential (\ref{EYMM}).
In terms of this potential, the partitiion function is
\bea
Z=\int [dM]e^{N{\rm Tr} V(M)}
\eea
We will work entirely in the small phase space where all couplings are set to zero.
Thus, we reduce (\ref{EYMM}) to
\bea
   {\rm Tr}V(M) = -2 {\rm Tr} (M{\rm ln}M -M)
\eea
Starting from the Schwinger-Dyson equation
\bea
  0=\int [dM]{d\over dM_{ij}}\left( (M^{2n_1-1})_{ij}\prod_{j=2}^k{\rm Tr}(M^{2n_j})e^{N{\rm Tr} V(M)}\right)
\eea
proceeding as explained above for the Gaussian matrix model,
we obtain the following recursion relation
\bea
   \gamma(\bar n_1 , n_2,\cdots,n_k)=
\sum_P\sum_{j=0}^{n_1-1}\gamma(j,P)\gamma(n_1-j-1,\tilde P)\cr
+\sum_{r=2}^k 2n_r \gamma (n_1+n_r-1,S'_r)
\eea
Although we did not use it for the class of correlators we focused on, note that the following recursion relation is also
easily obtained
\bea
   \gamma(\bar n_1 , \bar n_2,\cdots, n_k)=
\sum_P\sum_{j=0}^{n_1-1}\gamma(j,P)\gamma(n_1-j-1,\tilde P)\cr
+n_2 \gamma (\overline{n_1+n_2-1},S'_r)
+ \gamma (n_1+n_2-1,S'_r)
+\sum_{r=3}^k n_r \gamma (n_1+n_r-1,S'_r)
\eea
Further generalizations are straight forward.

\section{Topological A-model String Theory Correlators}\label{Amodel}

In this Appendix we summarize the recursion relations needed to compute the correlators used in this study.
Our discussion is rather brief and the reader is referred 
to \cite{Eguchi:1995er,Eguchi:1994in,Eguchi:1996tg,Gopakumar:2011ev,rajesh2,Hori:1994nb}
 for background.
Some of the correlators we use have been computed 
in \cite{Eguchi:1995er,Eguchi:1994in,Eguchi:1996tg,Gopakumar:2011ev,rajesh2}.
The new correlators that we use, that are not quoted in the literature, are derived in this Appendix.

There is a primary field for each cohomology class of the target manifold in the topological A-model string theory.
For the A-model on ${\bf P}^1$ there are thus two primaries, the puncture $P$ and the K\"ahler class $Q$.
The observables of the theory are these primaries and their gravitational descendents $\sigma_n(P)$, $\sigma_n(Q)$,
$n\ge 1$. 
By setting $O_1=P$ and $O_2=Q$, we describe the complete set of observables using the notation $O_\alpha$
and $\sigma_n (O_\alpha)$.
The $U(1)$ charges of the primary fields are $q_1=0$ and $q_2=1$.
We raise and lower the indices of the primary fields using the metric $\eta_{11}=\eta_{22}=0$ and
$\eta_{12}=\eta_{21}=1$.

The correlators of the theory are determined by solving recursion relations, which express $n$-point correlators in terms
of lower point correlators.
The partition function of the string theory contains a coupling constant for each observable in the theory.
One distinguishes between recursion relations that hold in the large phase space (when all couplings are turned on) and
recursion relations that hold only after certain couplings are set to zero.
When all the couplings are set to zero the genus $g$ correlation functions 
\bea
\langle \sigma_{n_1}(O_{\alpha_1})\cdots \sigma_{n_s}(O_{\alpha_s})\rangle_g
\eea
only recieve contributions from holomorphic maps of degree $d$ with
\bea
   2d + 2(g-1)=\sum_{i=1}^s (n_i+q_{\alpha_i}-1)
\eea
This is a consequence of the ghost number conservation law.
When we want to indicate the contribution to a correlator from maps of a specific degree $d$ at genus $g$, we will write
$\langle\cdots\rangle_{g,d}$.

The Eguchi-Hori-Yang relation (below $O_3=0$)
\bea
d^2 \langle\sigma_n (O_\alpha)\rangle_{0,d} = -2nd\langle\sigma_{n-1}(O_{\alpha+1})\rangle_{0,d}+\sum_{\bar d}
n\, \bar d^2\langle\sigma_{n-1}(O_\alpha )O_\beta\rangle_{0,d-\bar d}\langle O^\beta\rangle_{0,\bar d}
\label{EHY}
\eea   
holds in the large phase space.
Equation (B.5) of \cite{rajesh2} spells this out nicely.
Setting all couplings to zero and using $\langle P\rangle_{0,d} = \langle Q\rangle_{0,d} = 0$ except $\langle Q\rangle_{0,1}=1$,
the relation (\ref{EHY}) implies
\bea
  \langle\sigma_{2m}(Q)\rangle_{0,d}=\delta_{d,m+1}{(2m)!\over (m+1)!(m+1)!}
\eea
\bea
  \langle\sigma_{2m+1}(P)\rangle_{0,d}=-2c_{m+1}\delta_{d,m+1}{(2m+1)!\over (m+1)!(m+1)!}
\eea

Next, using the puncture equation
\bea
\langle P\prod_{i=1}^s\sigma_{n_i}(O_{\alpha_i})\rangle_0
=\sum_{i=1}^s n_i
\langle \sigma_{n_1}(O_{\alpha_1})\cdots\sigma_{n_i-1}(O_{\alpha_i})\cdots \sigma_{n_s}(O_{\alpha_s})\rangle
\eea
Hori's relation\cite{Hori:1994nb}
\bea
\langle Q\prod_{i=1}^s\sigma_{n_i}(O_{\alpha_i})\rangle_{0,d}
=d \langle \prod_{i=1}^s\sigma_{n_i}(O_{\alpha_i})\rangle_{0,d}\cr
+\sum_{i=1}^s n_i
\langle \sigma_{n_1}(O_{\alpha_1})\cdots\sigma_{n_i-1}(O_{\alpha_i+1})\cdots \sigma_{n_s}(O_{\alpha_s})\rangle_{0,d}
\label{Hori}
\eea
as well as the recursion relation between two point correlators, 
\bea
\langle\sigma_n(O_\alpha)\sigma_m(O_\beta)\rangle &=&
{2 mn \delta_{\phi\sigma}\eta^{\delta\phi}\eta^{\gamma\sigma}\over n+m+q_\alpha+q_\beta}
\langle\sigma_{n-1}(O_\alpha)O_\gamma\rangle
\langle\sigma_{m-1}(O_\beta)O_\delta\rangle\cr
&-&{2n \over n+m+q_\alpha+q_\beta}\langle\sigma_{n-1}(O_{\alpha+1})\sigma_m(O_\beta)\rangle\cr
&-&{2m \over n+m+q_\alpha+q_\beta}\langle\sigma_{n}(O_{\alpha})\sigma_{m-1}(O_{\beta+1})\rangle
\label{tpr}
\eea
we find
\bea
   \langle\sigma_{2m_1}(Q)\sigma_{2m_2}(Q)\rangle_0 
        = {1\over m_1+m_2+1}{(2m_1)!\over m_1! m_1!}{(2m_2)!\over m_2! m_2!}
\eea
\bea
   \langle\sigma_{2m_1-1}(Q)\sigma_{2m_2-1}(Q)\rangle_0 
        = {1\over 4(m_1+m_2)}{(2m_1)!\over m_1! m_1!}{(2m_2)!\over m_2! m_2!}
\eea
\bea
   \langle\sigma_{2m_1}(Q)\sigma_{2m_2+1}(P)\rangle_0 
        = {1\over m_1+m_2+1}{(2m_1)!\over m_1! m_1!}{(2m_2+1)!\over m_2! m_2!}\times\cr
         \times\left[-2c_{m_2}-{1\over m_1+m_2+1}\right]  \label{ts5}
\eea
\bea
   \langle\sigma_{2m_1+1}(Q)\sigma_{2m_2}(P)\rangle_0 
        = {1\over 2(m_1+m_2+1)}{(2m_1+2)!\over (m_1+1)! (m_1+1)!}{(2m_2)!\over m_2! (m_2-1)!}\times\cr
        \times \left[-c_{m_2}-c_{m_2-1}-{1\over m_1+m_2+1}\right] 
\eea
\bea
   \langle\sigma_{2m_1}(P)\sigma_{2m_2}(P)\rangle_0 
        = {2\over (m_1+m_2)^3}{(2m_1)!\over m_1! m_1!}{(2m_2)!\over m_2! m_2!}\cr
            \times (m_1m_2-(m_1+m_2) (m_2^2 c_{m_2}+m_1^2  c_{m_1})
            +2m_1m_2(m_1+m_2)^2 c_{m_1}c_{m_2})
\eea
Higher point correlators are now easily obtained by making use of the topological recursion relation
\bea
   \langle\sigma_n (O_\gamma)XY\rangle_0 = n\langle\sigma_{n-1}(O_\gamma)O_\alpha\rangle_0\eta^{\alpha\beta}
                  \langle O_\beta XY\rangle_{0}
\eea
which holds in the large phase space.
Application of the topological recursion relation now gives
\bea
   \langle \sigma_{2k_1-1}(Q)\sigma_{2k_2-1}(Q)\sigma_{2k_3}(Q)\rangle ={1\over 4}
{(2k_1)!\over (k_1!)^2}{(2k_2)!\over (k_2!)^2}{(2k_3)!\over (k_3!)^2}
\eea
\bea
   \langle \sigma_{2k_1-1}(Q)\sigma_{2k_2-1}(Q)\sigma_{2k_3+1}(P)\rangle =-{2c_{k_3}\over 4}
{(2k_1)!\over (k_1!)^2}{(2k_2)!\over (k_2!)^2}{(2k_3+1)!\over (k_3!)^2}
\eea
\bea
\langle\sigma_{2n}(P)\sigma_{2k_1-1}(Q)\sigma_{2k_2}(Q)\rangle =
{(2k_1)!\over (k_1!)^2}{(2k_2)!\over (k_2!)^2}{(2n-1)!\over ((n-1)!)^2}\left(-2c_n+{1\over n}\right)
\eea
Finally, we have also computed the higher point correlators
\bea
\langle \prod_{i=1}^n \sigma_{2k_i}(Q)\sigma_{2k+1}(P)\rangle =
(d+1)^{n-2}\prod_{i=1}^n {(2k_i)!\over k_i!k_i!}{(2k+1)!\over k!k!}
\left(-2c_{k}+{n-2\over d+1}\right)
\eea
\bea
\langle \sigma_{2k_1-1}(Q)\sigma_{2k_2-1}(Q)\prod_{i=3}^n \sigma_{2k_i}(Q)\sigma_{2k+1}(P)\rangle =
{d^{n-2}\over 4}\prod_{i=1}^n {(2k_i)!\over k_i!k_i!}{(2k+1)!\over k!k!}
\left(-2c_{k}+{n-2\over d}\right)\cr
\eea
where $d=\sum_i k_i+k$.
 
\section{Correlators in the Gaussian Matrix Model by orthogonal polynomials}

By making use of orthogonal polynomials it is possible to compute the connected correlators we have studied in this article.
If we focus on the concrete example of the Gaussian matrix model, the relevant orthogonal polynomials are the Hermite
polynomials.
In this case we have managed to explicitely verify all the integrals we have used by using Mathematica and also by
direct numerical evaluation.
In this way, there are a number of simple checks we can carry out which directly confirm the solutions to our recursion relations.
This has all been done without using analytic continuation to compute any of the integrals appearing\footnote{For these
examples, we are thus confirming that the integrals we need to compute can reliably be computed using analytic continuation.}.
Of course, establishing the validity of analytic continuation immediately confirms our solutions of the recursion relations, in
general.

For the Gaussian model we can rewrite the partition function in terms of Hermite polynomials
\bea
  Z&=&\int [dM]_{N\times N}e^{-{N\over 2}{\rm Tr}(M^2)}\cr
    &=&{\cal N}\int\prod_{i=1}^N d\lambda_i  \left({\rm det}\left[ H_k\left(\sqrt{N}\lambda_l\right)\right]\right)^2
            e^{-{N\over 2}\sum_{k=1}^N\lambda_k^2}\label{spf}
\eea
normalized using so-called ``probabilists'' conventions
\bea
H_0(x)&=&1\qquad H_1(x)=x\qquad H_2(x)=x^2-1\cr
H_3(x)&=&x^3-3x\qquad H_4(x)=x^4-6x^2+3\quad\cdots
\eea
We choose ${\cal N}$ in (\ref{spf}) such that $Z=1$.
Using the explicit form of these polynomials and standard techniques, its a trivial matter to verify that
\bea
   \int {d x\over\sqrt{2\pi}}e^{-{1\over 2}x^2}H_m(x)H_n(x)   =   \delta_{mn}\, m!
\eea
\bea
   xH_m(x)=H_{m+1}(x)+mH_{m-1}(x)
\eea
and hence that
\bea
   Z={\cal N}\left({2\pi\over N}\right)^{N\over 2}N!\prod_{l=1}^N l!\qquad\Rightarrow\qquad
   {\cal N}={1\over \left({2\pi\over N}\right)^{N\over 2}N!\prod_{l=1}^N l!}
\eea
A very useful formula, obtained from this recursion relation at large $N$, implies that
\bea
  x^{2k}H_{m}(x)=\cdots + {(2k)!\over (a)!(2k-a)!} m^a H_{m+2k-2a}(x)+\cdots
\eea
We will use this in performing integrals like (\ref{corrtest}) below.
Large N is used to assume that $m\gg 1$, which is true for very nearly all the terms in  (\ref{corrtest}).
This can be used to compute, for example,
\bea
   \int {d x\over\sqrt{2\pi}}e^{-{1\over 2}x^2}H_{2m}(x)H_{2n}(x) x^{2q}
  ={(2m)! (2q)!\over (m-n+q)!(q-m+n)!} (2m)^{m-n+q}\label{nolog1}
\eea
\bea
   \int {d x\over\sqrt{2\pi}}e^{-{1\over 2}x^2}H_{2m+1}(x)H_{2n+1}(x) x^{2q}
    ={(2m+1)!(2q)!\over (m-n+q)!(q-m+n)!} (2m+1)^{m-n+q}\label{nolog2}
\eea
which we will find extremely useful in what follows.
For example, we immediately find
\bea
{\cal N}\int\prod_{i=1}^N d\lambda_i \left({\rm det}H_i(\sqrt{N}\lambda_j)\right)^2
\,\sum_{l=1}^N \lambda^{2k}_l\, e^{-{N\over 2}\sum_{p=1}^N\lambda_p^2}
=\sum_{m=1}^N {(2k)!\over k! k!}\left({m\over N}\right)^k
\label{corrtest}
\eea
The last sum above can be evaluated at large $N$ by trading it for an integral. 
To do this replace ${m\over N}\to x$ and integrate over $x$ from 0 to 1.
In this way we obtain
\bea
{\cal N}\int\prod_{i=1}^N d\lambda_i \left({\rm det}H_i(\sqrt{N}\lambda_j)\right)^2
\,\sum_{l=1}^N \lambda^{2p}_l\, e^{-{N\over 2}\sum_{k=1}^N\lambda_k^2}
&=& N \int_0^1\, dx\, {(2k)!\over k! k!}x^k\cr
&=&N\, {(2k)!\over (k+1)! k!}
\eea
Of course, this reproduces the correct large $N$ result for the correlator
\bea
   \langle {\rm Tr}(M^{2k})\rangle = {(2k)!\over (k+1)! k!}N
\eea
We will also need
\bea
&& \int {d x\over\sqrt{2\pi}}e^{-{1\over 2}x^2}H_{2m}(x)H_{2n}(x) x^{2q} {\rm ln}(x^2)
={(2m)! (2q)!  (2m)^{m-n+q}\over (m-n+q)!(q-m+n)!}\times\cr
&& \times
\left({\rm ln}(2m)+2\sum_{j=1}^{2q}{1\over j}-\sum_{j=1}^{q+m-n}{1\over j}-\sum_{j=1}^{q-m+n}{1\over j}\right)
\label{withlog1}
\eea
\bea
&&\int {d x\over\sqrt{2\pi}}e^{-{1\over 2}x^2}H_{2m+1}(x)H_{2n+1}(x) x^{2q} {\rm ln}(x^2)
={(2m+1)!(2q)!(2m+1)^{m-n+q}\over (m-n+q)!(q-m+n)!}\times\cr
&&\times\label{withlog2}
\left({\rm ln}(2m+1)+2\sum_{j=1}^{2q}{1\over j}-\sum_{j=1}^{q+m-n}{1\over j}-\sum_{j=1}^{q-m+n}{1\over j}\right)
\eea
These formulas were obtained, for small values of $q$, using Mathematica.
The general $q$ values are then obtained by applying the recursion relations which follow from
\bea
   \int {d x\over\sqrt{2\pi}}{d\over dx}\left( e^{-{1\over 2}x^2}H_{m}(x)H_{n}(x) x^{q} {\rm ln}(x^2)\right)=0
\eea
Finally, we drop subleading terms at large $N$, which gives (\ref{withlog1}) and (\ref{withlog2}).
Notice that (\ref{withlog1}) and (\ref{withlog2}) can be obtained from (\ref{nolog1}) and (\ref{nolog2}) by analytic continuation.
This is basically what we are checking - that the analytic continuation gives the correct result.

\end{appendix}

\end{document}